\documentclass[letterpaper, 12 pt]{article}
\usepackage[a4paper, total={6in, 8in}]{geometry}
\usepackage{graphicx} 
\usepackage{amsmath}
\usepackage{graphicx}
\usepackage{cancel}
\usepackage{lipsum}
\usepackage{cite}
\usepackage{physics}
\usepackage{amsmath,amssymb}
\usepackage{dirtytalk}

\title{Near-extremal dumb holes and some aspects of the Hawking effect}
\author{ Akshat Pandey\footnote{apandey.physics@gmail.com} \\
\normalsize Department of Physics and Astronomy, KU Leuven, \\ \normalsize Celestijnenlaan 200d, Leuven B-3001, Belgium.}

\date{}
\begin{document}

\maketitle

\begin{abstract}
We propose novel non-relativistic fluid analogue models, that is dumb hole models,  for extremal and near-extremal black holes. Further we study the back-reaction effects of analogue Hawking radiation emitted from these dumb holes. We discuss and quantify the reduction in the background fluid velocity caused by radiation of Hawking phonons. In doing so, we speculate on the existence of an emergent Hawking force which leads to the reduction in the background fluid velocity and which is produced as a consequence of phonon emission. In addition to the analogue gravity literature, our results might be of relevance to black hole pedagogy.
\end{abstract}

\section{Introduction}

In 1981, Unruh discovered that a transsonic fluid flow leads to a thermal emission of quanta of sound much like a black hole horizon leads to Hawking radiation \cite{unruh}, hence the name ``dumb holes". Since then, dumb holes or acoustic black holes have become an integral part of the analogue gravity literature \cite{review, artificial bh, new review, nr2}. 

Like black holes, these acoustic black holes have horizons, and these horizons have a corresponding analogue Hawking temperature. However, there are some, rather clear limits to this acoustic analogy to black holes \textbf{---} they can reproduce only the kinematics of General Relativity (GR) or, (Quantum)  fields in curved spacetimes to be more precise, and not the dynamics of GR (Einstein equations) \cite{visserprl}. Because of this one cannot define quantities analogous to the black hole energy, or, Bekenstein entropy, and there exists no notion of an acoustic analogue to the laws of black hole thermodynamics \cite{visser}.

Despite these limits on the analogue model, there are some reasonable questions to ask, for example about back reaction effects; black holes upon radiating lose mass, what is the corresponding effect in dumb holes? There is an interesting paper by Patrick \textit{et al.} where they have a more general version of this question in a draining vortex flow \cite{backreaction}. This question, at least in principle has a well-defined answer \textbf{---} since it is the inhomogenous fluid flow that gives rise to the curvature of the acoustic spacetime and fluid at rest corresponds to the acoustic version of Minkowski spacetime \cite{IJMPA}, therefore radiation should result in a decrease in the speed of the fluid \cite{essential}. However, because of the limits mentioned above, there is very little quantitaive discussion of this effect. In the present paper we shall take a complementary perspective to \cite{backreaction}, we will try to start with a black hole first approach, think about the analogy with transsonic fluid flows and see how far we can extend the analogy.

In this paper, we shall see that asking a simple question \textbf{---} ``can we construct an acoustic analogue of an extremal black hole?" leads us to think about the above effect in a novel and, somewhat, quantitative way. Once we have an extremal dumb hole, we shall see that adding a small speed to the fluid in this metric would give us a near-extremal counterpart. Analysing the emission of phonons in this near extremal dumb hole spacetime, where the departure from extremality is due to the extra speed suggests that indeed it leads analogue Hawking radiation indeed leads to a decrease in the speed of the fluid. Since a change in speed is associated with a force, we shall  in section 3 speculate on the idea that due to radiation the fluid experiences an emergent force which in turn reduces the speed of the fluid.

This paper is organised as follows. In section 2, we shall construct the acoustic extremal black hole, we shall check that it is indeed an extremal solution. We shall then use this solution to construct an acoustic analogue for near extremal black holes. In section 3 we shall study the Hawking effect in this near extremal dumb hole, analyse the change in the fluid speed and note its consequences. We shall end with a summary and some discussion in section 4.

\section{Extremal and Near-extremal dumb holes}

We start with the acoustic metric in (1+1) dimensions

\begin{equation}
    ds^2 = - (c_s^2 - v_0^2) dt^2 + 2 v_0dx dt + dx^2 
\end{equation}

Here $c_s$ is the speed of sound and $v_0(x,t)$ is the background fluid speed. The above line element can be thought of as the (3+1) dimensional line element with the fluid velocity profile being non-zero only in the $x$ direction \cite{GnC}. Further, we can think of the fluid flowing through a cylinder and we are focusing only on the region around the cylindrical axis, ignoring the region close to the walls and thus any interactions with the walls and transverse flows \cite{dave}. 

As mentioned in the introduction it is the velocity field $v_0 (x, t)$ that gives rise to ``curvature" in acoustic spacetimes; different choices of velocity $v_0$ profiles lead to different geometries. 

In order to get an extremal black hole, we make the following choice for the velocity field 

\begin{equation}
    v_0(x, t) = v_0(x) = c_s (\cos x) 
\end{equation}

Additionally,  the flow is from positive to negative $x$ (this would be like the choice of convention for horizons, once we introduce non-extremality). Physically, this can be thought of as twisting the above mentioned cylinder into an oscillator such that the fluid flows from positive to negative $x$, its speed at each spatial point $x$ determined by the cosine function. 

With this $v_0$, the line element becomes

\begin{equation}
    ds^2= - (c_s^2 - c_s^2 \cos^2x) dt^2 + 2 \cos xdx dt + dx^2
\end{equation}

This choice for $v_0$ ensures that at $x=0$, we get $v_0 = c_s$ i.e. an extremal horizon.

We can compare the above extremal dumb hole solution with the extremal Reissner–Nordstr\"{o}m (RN) solution. First, we make the following time coordinate transformation to equation (3) \cite{painleve, parikh}

\begin{equation}
    d t_s = d t + \frac{v_0}{1-v_0^2}\mathrm{d}x = d t + \frac{\cos x}{\sin^2x}\mathrm{d}x.
\end{equation}

In these coordinates, the line element becomes

\begin{equation}
    d s^2 = - c_s^2(\sin ^2x)dt_s^2 + \frac{dx^2}{\sin ^2x}
\end{equation}

The extremal RN solution is \cite{carroll}

\begin{equation}
    d s^2 = - \left(1-\frac{M}{r}\right)^2 dt^2 + \frac{d r^2}{\left(1-\frac{M}{r}\right)^2} + r^2 d \Omega^2
\end{equation}

We see that the extremal dumb hole solution is indeed like the extremal RN solution, in the sense that the metric coefficients are total squares and thus always positive.

The Hawking temperature of extremal black holes is known to vanish. As another quick check, we can calculate the Hawking temperature (or simply the surface gravity) of the extremal dumb hole solution above. The surface gravity $\kappa$ turns out to be \cite{acoustic1, acoustic2}

\begin{equation}
    \kappa = -c_s \frac{dv_0}{dx} \bigg|_{x=0} = c_s \sin 0 = 0 
\end{equation}

Now that we have modelled an extremal solution, we would like to see how we can construct a near-extremal solution from it. We start with the $v_0$ corresponding to the extremal dumb hole and add a homogenous (in $x$) infinitesimal speed such that

\begin{equation}
    v_0 = c_s(\cos x + \beta)
\end{equation}

Here $\beta$ is an infinitesimal constant and can be thought of as the non-extremality parameter. Physically this can be thought of as adding a (spatially) constant infinitesimal speed $\delta v_0= c_s \beta$ to the speed profile for the extremal case. Additionally, we would assume that the speed profile mentioned in equation (8) is restricted only to a small region around $x=0$. We shall return to this assumption in section 3.  

This implies that the horizons are located at $x = \pm \theta$ such that 

\begin{equation}
    \cos \theta + \beta = 1
\end{equation}

Here $|\theta| > 0$. There are two horizons, the one at $x= \theta$ corresponds to an analogous black hole horizon, while the $x= - \theta$ corresponds to an analogous white hole horizon. We can see that this is due to the choice of the direction of fluid flow; from positive to negative $x$. We shall restrict our attention to the black hole horizon, $x= \theta$.

In terms of the non-extremality parameter, this horizon is located at
\begin{equation}
    \theta = \cos ^{-1} (1 - \beta)
\end{equation}

Given $\beta$ is small, we can see that $x=\theta$ would be close to $x=0$. Further, plugging $\beta=0$ implies the horizon is at $x=0$ i.e. the extremal case. 

As a quick check that equation (9) makes sense as the velocity profile for a near-extremal dumb hole, we calculate the surface gravity

\begin{equation}
    \kappa = -c_s \frac{dv_0}{dx} \bigg|_{x=\theta}=  c_s \sin \theta |_{\theta = \cos^{-1}(1 - \beta)}
\end{equation}

Since, $\theta$ is non-zero we can see that the surface gravity and consequently the Hawking temperature $T_H$ are non zero, where the latter is equal to

\begin{equation}
    T_H = \frac{\kappa}{2 \pi} = \frac{ \sin (\cos ^{-1} (1 - \beta)) c_s}{2 \pi}
\end{equation}

Therefore, we have constructed an analogue model for a near-extremal black hole.

\section{A Hawking force?}

We now return to the question posed in the introduction namely can we quantify the reduction in the fluid speed due to phonon radiation? There is one immediate conceptual issue. Heuristically, in the case of black hole evaporation, (Schwarszchild, say), the emission of one Hawking quanta of energy $\omega$ leads in a reduction of the black hole mass from $M$ to $M- \omega$ (in the appropriate units). Unlike mass, which is a global charge of the spacetime, in the acoustic analogue it is not clear if the reduction in the fluid velocity would be locally at the horizon \footnote{Expanding on what we mentioned in the introduction, the immediate issue one runs into is the absence of a consistent analogue for energy and entropy of black holes, therefore the absence of a consistent analogue for the laws of black hole thermodynamics. This is quite reasonable as what the acoustic analogue truly provides us with is an analogue for a horizon, and not for the global black hole geometry.} or would it require the details of the Bose-Einstein Condensate (BEC), in whose hydrodynamic limit we are studying these quantised sound waves (phonons) \cite{BEC1, BEC2, BEC3}.

In this present paper, working with near-extremal dumb holes, we circumvent the problem by focussing on the energy of the radiated phonons \textbf{---} the absence of analogous dynamics prevents us from defining energy to the dumb hole, however the energy radiated as phonons is a perfectly well-defined notion \footnote{Given a temperature and energy one might be interested in defining an entropy for the radiated hawking phonons. It would then be interesting to see if it at all is related to the von Neumann (or thermodynamic) entropy of the phonons. Although somewhat speculative, we believe this could be an interesting problem if made precise}. 

It is only the infinitesimal $\delta v_0 = c_s \beta$ that leads to radiation therefore the infinitesimal energy $\epsilon$ (say), radiated as Hawking phonons have to be $\delta E \equiv \delta E(v_0)= \epsilon$ . Note, the above definition has an implicit assumption \textbf{---} radiation of phonons leads to a reduction in fluid speed which is uniform throught the fluid, atleast approximately. This is why while defining the speed profile for the near-extremal case we mentioned that we assume the fluid is restricted to a small region near $x=0$. Being restricted to a small spatial region, with the change in the energy and change in the fluid speed both being infinitesimal, we can see that the uniform (infinitesimal) reduction in the fluid speed is a reasonable assumption. Additionally, we set the total mass of the fluid $m_{fl}$ in this small region to be equal to $m_{fl} = \rho_0 V =1$. Here $\rho$ is the background fluid density and $V$ is the volume of the fluid, setting it to $1$ is simply for convinience.

A useful quantity to calculate is the radiated power \cite{giddings}, which following from Stefan-Boltzmann law is

\begin{equation}
    \frac{d E}{d t} = A \sigma T^4
\end{equation}

Here, $A$ is the area of the emitting black body which in this case is the $x=\theta$ horizon area, $\sigma$ is the Stefan-Boltzmann constant and $T$ is the temperature which in this case is the Hawking temperature $T_H$. One could be more precise and include grey-body factors, however given that we are working in effectively (1+1) dimensions, we shall ignore grey-body corrections (if it were a pure (1+1) dimensional system, the grey-body factors would have been 0 \cite{harmark}).

Using the chain rule and approximating in terms of infinitesimals

\begin{equation}
     \frac{d E}{d t} = \frac{\delta E}{\delta v_0} \frac{dv_0}{dt} = A \sigma T_H^4
\end{equation}

Plugging in the result for $T$, we can write 

\begin{equation}
    \frac{d v_0}{dt} =  \frac{c^5_s \beta}{\epsilon}\sigma A \sin^4(\cos^{-1}(1- \beta)) 
\end{equation}

 Since we have imposed $m_{fl}=1$ we can define this as a force namely

\begin{equation}
    F_H = \frac{d v_0}{dt} =   \frac{c^5_s \beta}{\epsilon}\sigma A \sin^4(\cos^{-1}(1- \beta))
\end{equation}

Therefore, we have found at least for the infinitesimal case, an expression for force exerted on the fluid due to analogue hawking radiation, in terms of the fluid velocity and the energy radiated.

\section{Summary and discussion}

In this paper, first, we proposed new acoustic analogue models for extremal and near-extremal black holes. We did this by proposing a fluid velocity profile proportional to the cosine of the distance for the extremal case and small corrections to the cosine for the near-extremal case. The outcome of these choices, these dumb hole solutions, are physically quite intuitive and straightforward. Further, in pedagogy, just how dumb holes serve as a useful analogy to introduce the physics of black holes \cite{dhpop}, we believe our proposed models would be useful as intuitive analogies for extremal and near-extremal black holes.

Second, in the context of the near-extremal dumb hole, we studied the back-reaction effects of the radiation of Hawking phonons, which intuitively should correspond to a reduction in the background fluid velocity. The near-extremal approximation, with the additional assumption of restricting our system to a small spatial region, allowed us to go around the limits of the acoustic analogue; we tried to quantify this reduction in fluid velocity. To this effect, we speculated on the existence of a retarding force which emerged as a consequence of analogue Hawking radiation.

While the former, the proposed analogue models, are at least on solid theoretical foundations, the latter i.e. our study of back-reaction and the proposal of the Hawking force is somewhat sketchy. Therefore we would like to end with two comments about this.

\begin{itemize}
    \item The derivation in section 3 is based on several assumptions, most central of all is the hydrodynamic limit \cite{BEC4}. A more bottom-up BEC based analysis might help understand if the derivation is at all valid. In case it is valid, it would still remain to see if it is simply an obvious corollary feature of artificial black holes or if it points to something truly interesting \cite{jeff}. 

    \item We defined an emergent force, one that seeks to extremalise the spacetime. This force is so straightforward that it barely deserves the name. However, thinking about the consequences Hawking radiation and black hole thermodynamics has, at least indirectly, led to several postulates of emergent forces, for instance Verlinde's proposal of gravity being an emergent force arising as a consequence of increase in entropy \cite{verlinde}.

\end{itemize}

\section*{Data Availability Statement}

No new data was produced in the preparation of this manuscript.

\end{document}